\documentclass[12pt]{article}
\usepackage{graphicx}

\newcommand\pubnumber{SNSN-323-63}
\newcommand\pubdate{\today}

\def\oxford{University of Oxford\\
Clarendon Laboratory, Parks Road, Oxford OX1 3PU, UK}
\def\ipmu{Kavli Institute for the Physics and Mathematics of the Universe\\
University of Tokyo, 5-1-5 Kashiwanoha, Kashiwa 277-8583, Japan}
\def\support{\footnote{for the T2K collaboration}}

\def\Title#1{\begin{center} {\Large #1 } \end{center}}
\def\Author#1{\begin{center}{ \sc #1} \end{center}}
\def\Address#1{\begin{center}{ \it #1} \end{center}}

\newcommand\pubblock{\rightline{\begin{tabular}{l} \pubnumber\\
         \pubdate  \end{tabular}}}
\newenvironment{Abstract}{\begin{quotation}  }{\end{quotation}}
\newenvironment{Presented}{\begin{quotation} \begin{center} 
             PRESENTED AT\end{center}\bigskip 
      \begin{center}\begin{large}}{\end{large}\end{center} \end{quotation}}

\def\beq{\begin{equation}}
\def\eeq#1{\label{#1}\end{equation}}
\def\eeqn{\end{equation}}

\def\beqa{\begin{eqnarray}}
\def\eeqa#1{\label{#1}\end{eqnarray}}
\def\eeqan{\end{eqnarray}}

\let\bar=\overbar

\def\Dslash{\not{\hbox{\kern-4pt $D$}}}
\def\dslash{\not{\hbox{\kern-2pt $\del$}}}

\def\msb{{\bar{\ssstyle M \kern -1pt S}}}

\newcommand{\upsub}[1]{\sb{\mathrm{#1}}}
\newcommand{\upsup}[1]{\sp{\mathrm{#1}}}

\begingroup\lccode`~=`_\lowercase{\endgroup\let~\upsub}
\begingroup\lccode`~=`^\lowercase{\endgroup\let~\upsup}

\AtBeginDocument{%
  \catcode`_=12 \catcode`^=12
  \mathcode`_="8000
  \mathcode`^="8000
}

\begin{document}
\begin{titlepage}
\pubblock

\vfill
\Title{Constraining the T2K Neutrino Flux Prediction with 2009 NA61/SHINE Replica-Target Data}
\vfill
\Author{ Tomislav Vladisavljevic\support}
\Address{\oxford \newline
\newline
\ipmu}
\vfill
\begin{Abstract}
Accurate modelling of the T2K neutrino flux is crucial for a better understanding of neutrino interactions at the near and far detectors. Most of T2K neutrinos are created through in-flight decays of unstable hadrons, produced by interactions of 31 GeV/c protons in a long graphite target (90 cm). External hadron production data is used for correcting the flux model. The analysis presented here uses a new NA61 dataset, collected in 2009 using the full length replica of the T2K target. The preliminary results suggest a reduction of the hadronic interaction component of the neutrino flux uncertainty by $\sim$50\%, to errors of less than 5\%.
\end{Abstract}
\vfill
\begin{Presented}
NuPhys2017, Prospects in Neutrino Physics\\
Barbican Centre, London, UK,  December 20-22, 2017
\end{Presented}
\vfill
\end{titlepage}
\def\thefootnote{\fnsymbol{footnote}}
\setcounter{footnote}{0}

\section{The T2K Neutrino Flux}

The T2K (Tokai-to-Kamioka) experiment \cite{Abe:2011ks} is a long baseline neutrino experiment located in Japan. The main goal of the experiment has now shifted from observing electron neutrino appearance $\nu_{\mu}\rightarrow\nu_e$ \cite{Abe:2011sj} to measuring CP violation in neutrino mixing. An intense muon (anti)neutrino beam $\nu_{\mu}$ ($\bar{\nu}_{\mu}$) is fired across Japan, from the village of Tokai, on the eastern coast of Japan, to the Super Kamiokande (SK) detector located in the mountains on the western side of the island, 295 km away.

The initial $\nu_{\mu}$ ($\bar{\nu}_{\mu}$) beam is produced inside the Japan Proton Accelerator Research Complex (J-PARC), where 31 GeV/c protons are aimed at a graphite target, and the pions and kaons resulting from this collision are guided into the decay volume using magnetic horns. Switching the horn polarity allows for the focusing of either positively or negatively charged hadrons, which respectively produce either neutrinos or anti-neutrinos through decays in the decay volume. T2K uses two near detectors, ND280 and INGRID, to measure the neutrino flux just after production point, and SK measures the neutrino flux 295 km downstream. Neutrino oscillation parameters can be extracted from the change in the composition of the neutrino flux between the near and far detectors.

\section{The NA61/SHINE Measurements for T2K}

\begin{figure}[htb]
\centering
\includegraphics[width=0.75\textwidth]{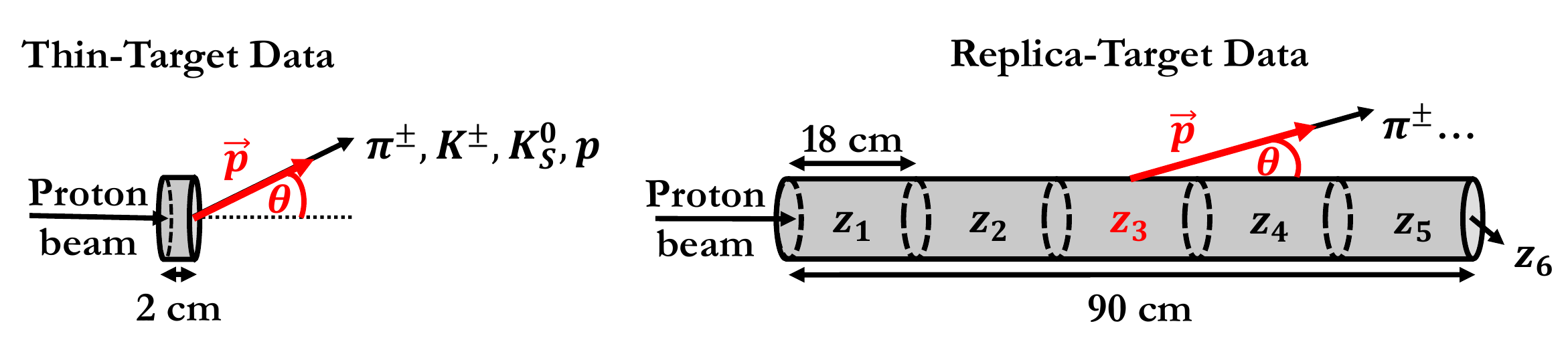}
\caption{The NA61 thin-target and replica-target configurations used for hadron production measurements for T2K.}
\label{fig:NA61TargetConfigurations}
\end{figure}

The NA61/SHINE (SPS Heavy Ion and Neutrino physics Experiment) \cite{Abgrall:2014xwa} is a fixed target experiment served by the H2 beam line of the CERN North Area. The experiment has been proposed in November 2006 and inherited many of its components from NA49. It is a multi purpose research facility providing precise hadron production measurements for long baseline neutrino experiments (T2K, NO$\nu$A, MINER$\nu$A etc.), used in reducing the unoscillated neutrino flux uncertainty. For T2K, NA61 measured the differential multiplicities (yields) of charged hadrons exiting from two distinct target configurations, the thin-target \cite{Abgrall:2015hmv} and the replica-target \cite{Abgrall:2016jif} (see Fig.~\ref{fig:NA61TargetConfigurations}). With the thin-target dataset, the measured hadronic yields are binned by the outgoing hadron momentum and angle $N^{NA61}_{thin}\left(p,\theta\right)$, whereas the replica-target multiplicity measurements are also binned based on the outgoing hadron's exiting position along the target $N^{NA61}_{replica}\left(p,\theta,z\right)$.

Presented here are the preliminary uncertainties associated with the NA61 2009 replica-target tuned T2K neutrino flux prediction. Currently, the T2K collaboration still relies on the NA61 2009 thin-target dataset to constrain the neutrino flux prediction. This dataset is appropriate for directly constraining $\sim$60\% of the neutrino flux which originates from primary interactions within the graphite target. The strength of the replica-target dataset lies in its ability to directly constrain both primary interactions and subsequent reinteractions within the target, thus accounting for $\sim$90\% of T2K neutrino flux. Due to limited statistics, the 2009 replica-target dataset only contains charged pion yields, so that thin-target data is still used for constraining the neutrino yield for other types of hadrons ($p,K^{\pm},K^0_S,K^0_L$), or for exiting pions outside the $\left(p,\theta\right)$ phase space of the replica-target measurements.

\section{Constraining the T2K Neutrino Flux Prediction}

T2K neutrino flux predictions \cite{Abe:2012av} rely on modelling interactions of primary protons incident on the graphite target, the propagation and interaction of subsequent hadrons resulting from this primary interaction, and the eventual decay of daughter hadrons into neutrinos. The simulation is driven by proton beam profile and horn current measurements, and based on a combination of FLUKA2011 \cite{Ferrari:2005zk}\cite{Bohlen:2014buj}, GEANT3 \cite{Brun:1994aa} and GCALOR \cite{Fasso:1993kr}. Associated with every simulated neutrino is a chain of hadronic interactions leading to its production. At every interaction point, the kinematic information for the incident and outgoing particle is stored, in addition to the target nucleon species and the distance travelled by particles through each of the simulated detector components. Finally, the nominal flux prediction gets constrained based on available hadron measurements, and the associated flux uncertainty evaluated. Weights are applied to every simulated neutrino event based on its hadronic history: every interaction in the ancestry chain is assigned a multiplicity weight, and every propagating parent hadron is assigned a hadron interaction length weight.
 
\begin{figure}[htb]
\begin{minipage}{6in}
\centering
\raisebox{-0.5\height}{\includegraphics[width=0.4\textwidth]{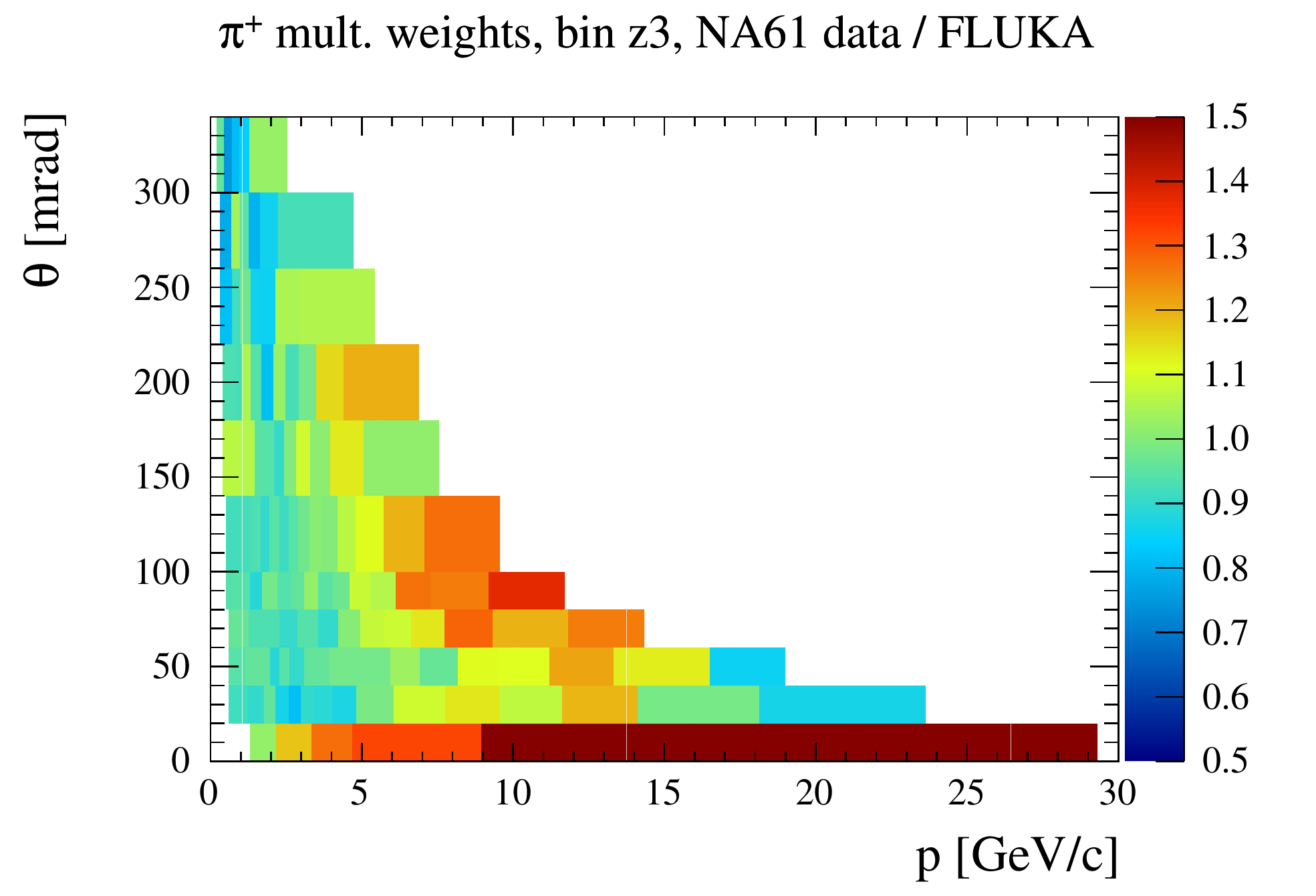}}
\raisebox{-0.5\height}{\includegraphics[width=0.4\textwidth]{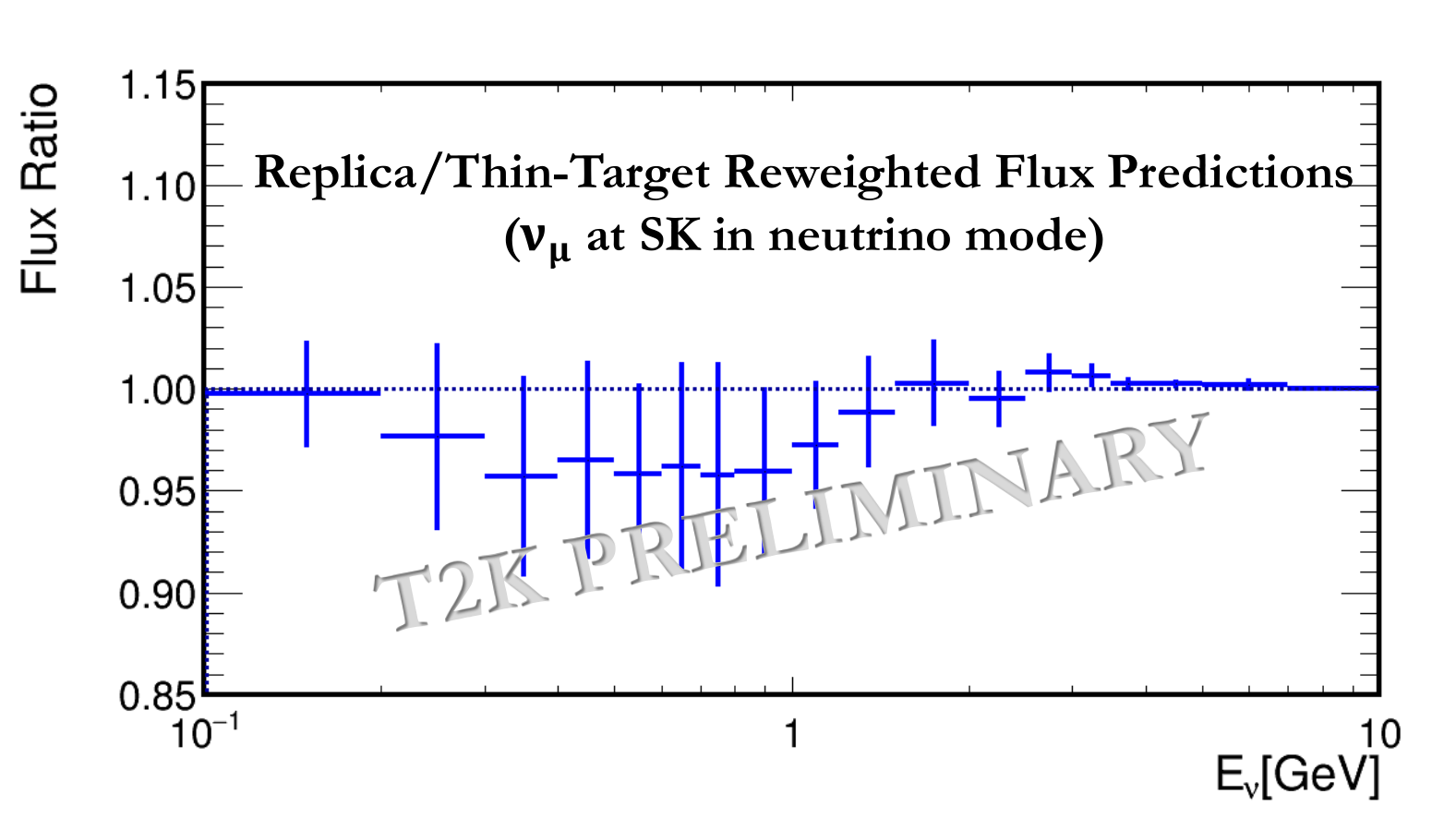}}
\end{minipage}
\caption{Replica-target positive pion $w(p,\theta,z_3)$ multiplicity weights (left side). Ratio of replica- and thin-target constrained flux predictions, with ratio errors propagated from the assigned production cross section uncertainty (right side).}
\label{fig:CompositeFigure}
\end{figure}

The multiplicity weight corrects the predicted neutrino yields based on the momentum $p$ and angle $\theta$ (measured with respect to the beam direction) of the produced ancestor hadron:
\begin{equation}
w_{thin}(p,\theta)=N^{NA61}\left(p,\theta\right)/N^{model}\left(p,\theta\right).
\end{equation}
If the ancestor hadron exiting from the target is a pion, it is sufficient to constrain only the outgoing pion rate with the corresponding replica-target weight $w_{replica}(p,\theta,z)$, instead of individually constraining all in-target interactions (see Fig.~\ref{fig:CompositeFigure} for representative replica multiplicity weights). Thin-target multiplicity weights are assigned to out-of-target interactions.

The interaction length weight corrects the predicted neutrino yield based on the distance $d$ travelled by propagating ancestor hadron through different materials before interacting:
\begin{equation}
w(p,d)=\frac{\sigma^{data}}{\sigma^{model}}\textrm{exp}\left(-\rho d\left(\sigma^{data}-\sigma^{model}\right)\right),
\end{equation}
where $\rho$ is the target material number density, $p$ is the propagating hadron's momentum and $\sigma\equiv\sigma(p)$ is the hadronic production cross section. The production cross section is assigned an uncertainty equal to the quasi-elastic cross section, to reflect the observed preference of the replica-target measurements for the proton production cross-section of $\sim$200 mb (see Fig.~\ref{fig:CompositeFigure}).

\noindent

\section{Results and conclusions}

\begin{figure}[htb]
\centering
\includegraphics[width=0.49\textwidth]{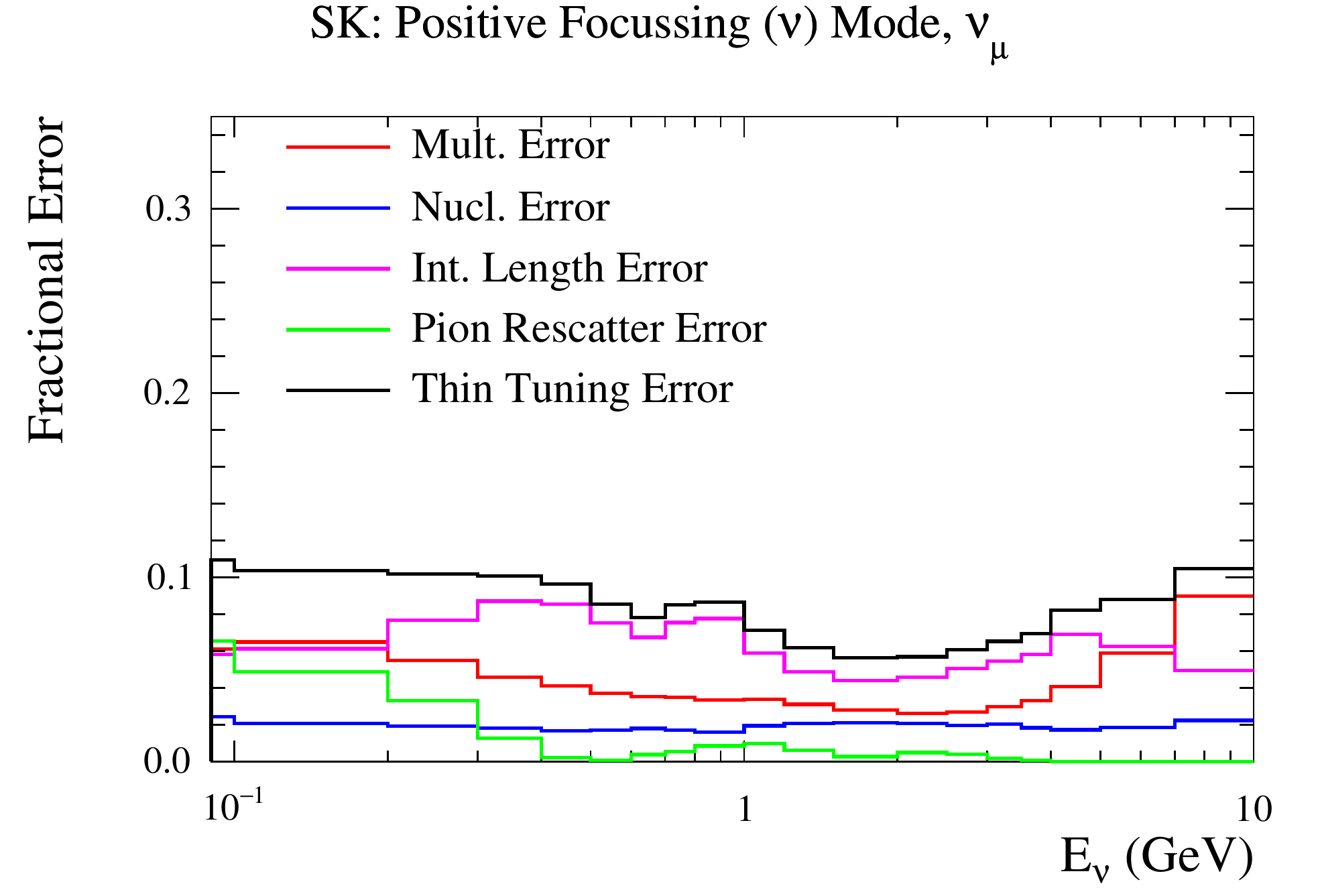}
\includegraphics[width=0.49\textwidth]{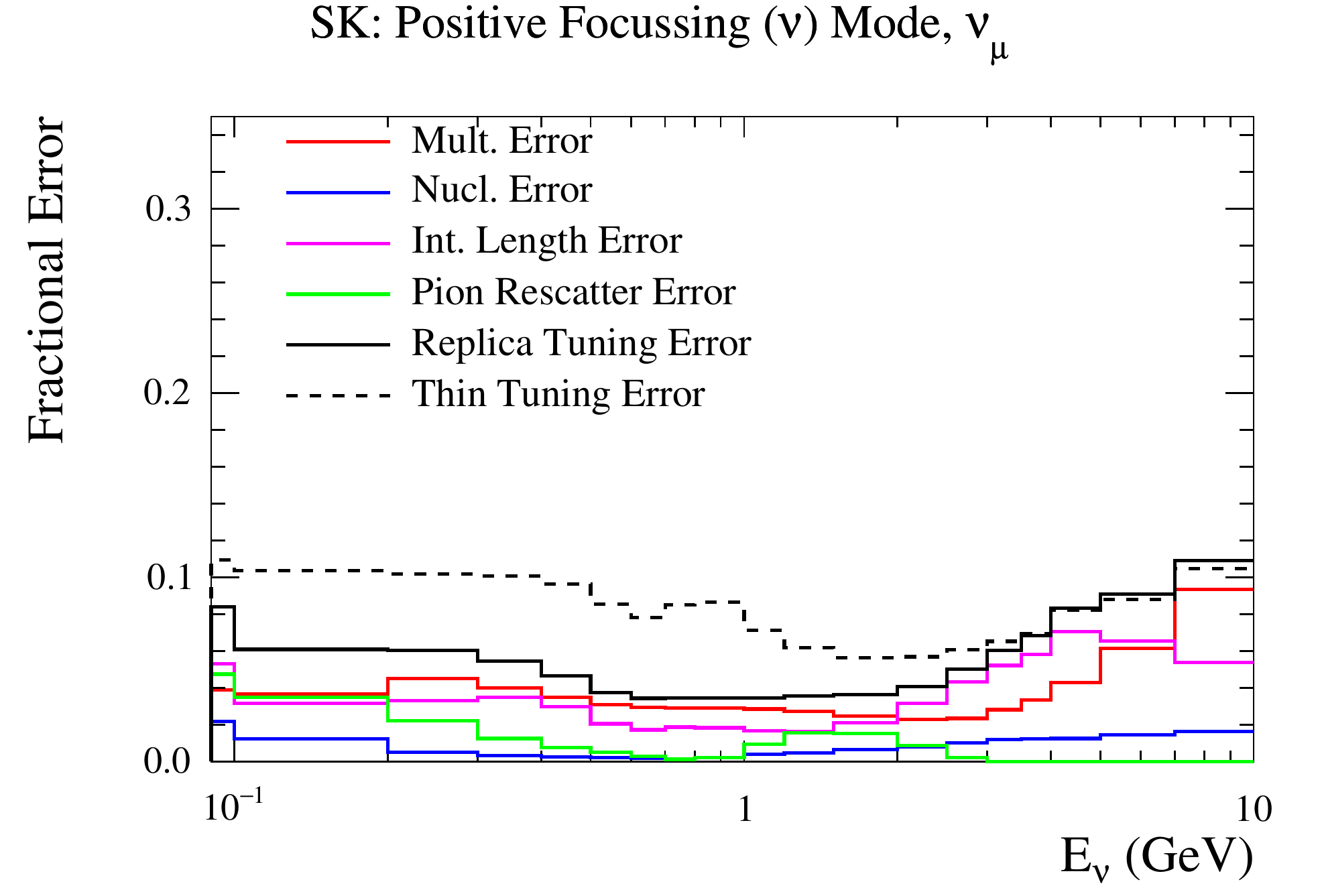}
\caption{The hadron interaction model uncertainties evaluated on the SK flux prediction. The uncertainties have been calculated for the flux constrained with either purely NA61 2009 thin-target data (left side), or using a combination of NA61 2009 thin-target and replica-target data (right side, denoted as the replica tuning error).}
\label{fig:FluxUncertainties}
\end{figure}

Neutrino flux uncertainty from the hadronic interaction model, as a function of neutrino energy, are shown in Fig.~\ref{fig:FluxUncertainties} (SK in neutrino mode). The pion rescattering error was estimated using HARP double differential pion cross section measurements \cite{Apollonio:2009bu}. The nuclear error comes from constraining secondary and tertiary baryon interactions using Feynman scaling and target nucleus scaling for extending the coverage of existing hadron production measurements. Around the T2K neutrino flux peak, the replica-tuned flux uncertainty is $\sim$50$\%$ smaller than the thin-tuned flux uncertainty. In particular, the hadron interaction length uncertainty, related to constraining the hadronic production cross section, is substantially reduced at lower neutrino energies. The hadronic multiplicity and pion rescattering uncertainties are also reduced.

The preliminary results suggest a 50\% reduction in the hadronic interaction component of the neutrino flux uncertainty, which could open up attractive prospects for the T2K neutrino cross section measurements programme.

\end{document}